\documentclass[12pt,preprint]{aastex}

\slugcomment{Draft version from \today}

\shorttitle{SMA jet/disk observations in IRAS\,18089-1732}
\shortauthors{Beuther et al.}

\begin{document}

\title{SMA outflow/disk studies in the massive star-forming region
IRAS\,18089-1732}


\author{H. Beuther\footnote{Harvard-Smithsonian Center for
       Astrophysics, 60 Garden Street, Cambridge, MA 02138},
       T.R. Hunter$^1$, Q. Zhang$^1$, T.K. Sridharan$^1$, J.-H.
       Zhao$^1$, P. Sollins$^1$, P.T.P. Ho$^{1,2}$,
       N. Ohashi\footnote{Academia Sinica Institute of Astronomy and
       Astrophysics, No.1, Roosevelt Rd, Sec. 4, Taipei 106, Taiwan,
       R.O.C.}, Y.N. Su$^2$, J. Lim, S.-Y. Liu$^2$}

\begin{abstract}
SMA observations of the massive star-forming region IRAS\,18089-1732
in the 1\,mm and 850\,$\mu$m band reveal outflow and disk signatures
in different molecular lines. The SiO(5--4) data show a collimated
outflow in the northern direction. In contrast, the HCOOCH$_3$(20--19)
line, which traces high-density gas, is confined to the very center of
the region and shows a velocity gradient across the core. The
HCOOCH$_3$ velocity gradient is not exactly perpendicular to the
outflow axis but between an assumed disk plane and the outflow
axis. We interpret these HCOOCH$_3$ features as originating from a
rotating disk that is influenced by the outflow and infall. Based on
the (sub-)mm continuum emission, the mass of the central core is
estimated to be around 38\,M$_{\odot}$. The dynamical mass derived
from the HCOOCH$_3$ data is 22\,M$_{\odot}$, of about the same order
as the core mass. Thus, the mass of the protostar/disk/envelope system
is dominated by its disk and envelope. The two frequency continuum
data of the core indicate a low dust opacity index $\beta \sim 1.2$ in
the outer part, decreasing to $\beta \sim 0.5$ on shorter spatial
scales.
\end{abstract}

\keywords{star: formation -- ISM: jets and outflows -- accretion disks
-- submillimeter -- techniques: interferometric}

\section{Introduction}

Unambiguous proof for disks in massive star formation is still
missing. Millimeter continuum observations suggest flattened
structures without providing velocity information (e.g.,
\citealt{shepherd2001}), and molecular line studies suggest rotational
motions but are often confused outflows and ambient gas (e.g.,
\citealt{zhang1998} and Beuther et al., this volume). Maser studies
show disk signatures in some cases but are mostly not
unambiguous as well (e.g., \citealt{churchwell2002}). The best
evidence yet for genuine disk emission comes from CH$_3$CN
observations in IRAS\,20126+4104 \citep{cesaroni1999}. In this case,
the velocity gradient defining the presence of the disk is aligned
perpendicular to the bipolar outflow, consistent with the common
disk/jet paradigm.  To further investigate possible disk emission and
its association with molecular jets, we used the Submillimeter Array
(SMA) to observe the jet tracer SiO(5--4) and the hot-core tracer
HCOOCH$_3$(20--19) in a massive star-forming region.

The source IRAS\,18089-1732 is a young High-Mass Protostellar Object
(HMPO) which has been studied in detail over recent years. The source
is part of a sample of 69 HMPOs selected mainly via infrared
color-color criteria and the absence of strong cm emission
\citep{sridha}. IRAS\,18089-1732 is approximately at a distance of
3.6\,kpc\footnote{The kinematic distance ambiguity is solved by
associating the region via the near- and mid-infrared surveys 2MASS
and MSX on larger scales with sources of known distance (Bontemps,
priv. comm.).} and its bolometric luminosity is about
$10^{4.5}$\,L$_{\odot}$ \citep{sridha}. Millimeter continuum
observations reveal a massive core $>2000$\,M$_{\odot}$ with H$_2$O
and CH$_3$OH maser emission, and a weak 1\,mJy source is detected at
3.6\,cm \citep{beuther2002a,beuther2002c}. As part of a single-dish CO
outflow study, wing emission indicative of molecular outflows was
detected but the CO map was too confused to define a bipolar outflow
\citep{beuther2002b}. During these observations, \citet{beuther2002b}
also observed SiO(2--1) at 3\,mm, and bipolar structure was detected
in the north-south direction. Furthermore, \citet{sridha} reported the
detection of the hot-core-tracing molecules CH$_3$CN and CH$_3$OH.
This letter focuses on the jet/disk observations and the (sub-)mm
continuum data. A description of the line forest observed
simultaneously is presented in an accompanying paper (Beuther et al.,
this volume).

\section{The Submillimeter Array (SMA)}
\label{obs}

IRAS\,18089-1732 was observed with the SMA\footnote{The Submillimeter
Array is a joint project between the Smithsonian Astrophysical
Observatory and the Academia Sinica Institute of Astronomy and
Astrophysics, and is funded by the Smithsonian Institution and the
Academia Sinica.} between May and July 2003 in two different
configurations with 3 to 5 antennas in the array.  The phase reference
center of the observations was R.A.[J2000] 18:11:51.4 and Dec.[J2000]
$-17$:31:28.5. The frequency was centered on the SiO(5--4) line at
217.105\,GHz, the HCOOCH$_3$(20--19) line at 216.967\,GHz could be
observed simultaneously in the same band. The HCOOCH$_3$ line consists
of 8 distinct components but is dominated by 4 of them which are
separated by 2.5\,MHz (corresponding to 3.5\,km s$^{-1}$). The
correlator bandwidth at that time was 1\,GHz with a frequency
resolution of 0.825\,MHz. We smoothed the SiO(5--4) data to a
spectral resolution of 3\,km\,s$^{-1}$ and the HCOOCH$_3$(20--19) data
to 2\,km\,s$^{-1}$ to increase the signal-to-noise ratio. The
continuum was constructed via averaging the line-free channels in the
upper side-band. The beam size at 217\,GHz was $2.7''\times 1.7''$ and
at 354 GHz $1.4'' \times 0.9''$. System temperatures in the
850\,$\mu$m band were between 300-900\,K and in the 1\,mm band around
200\,K. The continuum rms at 217\,GHz was $\sim 8$\,mJy and at
354\,GHz 40\,mJy. The flux calibration was estimated to be accurate to
$25\%$. For more details on the SMA, the observations and data
reduction, see the accompanying papers by Ho, Moran \& Lo and Beuther
et al. (this volume).

\section{Results}

\subsection{Dust continuum emission}

\begin{figure}[htb] 
\includegraphics[angle=-90,width=8.8cm]{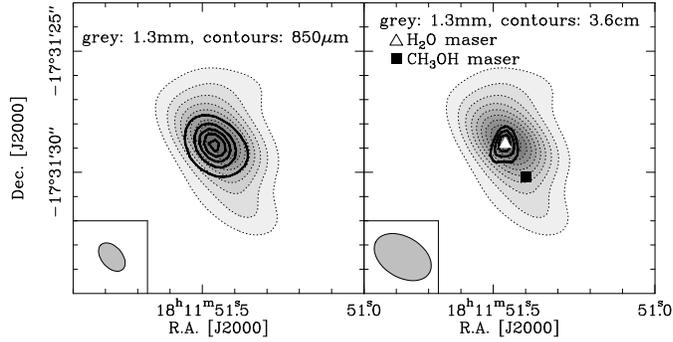}
\caption{The grey-scale (with dotted contours) shows the SMA 1.3\,mm
continuum emission in both panels. The left panel shows in heavy
contours the SMA 850\,$\mu$m continuum emission and the 850\,$\mu$m
synthesized beam. In the right panel we also present the cm continuum
emission and H$_2$O and CH$_3$OH maser position (taken from
\citealt{beuther2002c}), and the 1.3\,mm synthesized beam.
\label{continuum}}
\end{figure}

Figure \ref{continuum} compares the (sub-)mm continuum observations
and shows additional cm continuum and H$_2$O and CH$_3$OH maser data
\citep{beuther2002c}. Even in the highest-spatial-resolution data at
850\,$\mu$m, the dust emission remains singly peaked, i.e., it does
not split up into multiple sources as observed in other massive
star-forming regions, e.g., IRAS\,19410+2336 \citep{beuther2003a}.
Nevertheless, in our 1\,mm data we resolve elongated emission in the
south and north-west, which demonstrates that IRAS\,18089-1732 has a
compact mm core with extended halo emission
(Fig. \ref{continuum}). The halo emission is not seen in the
850\,$\mu$m observations because of the reduced sensitivity and
uv-coverage. While the weak 3.6\,cm peak and the H$_2$O maser position
coincide exactly with the (sub-)mm continuum peak, the CH$_3$OH maser
position is about $1.4''$ to the south. The latter could indicate that
there might be a second source at the position of the CH$_3$OH maser
which we cannot distinguish. Table \ref{para} shows the derived peak
and integrated fluxes ($S_{\rm{peak}}$ and $S_{\rm{int}}$) at 1\,mm
and 850\,$\mu$m. Comparing the SMA 1\,mm data with single-dish
observations of the region \citep{beuther2002a}, we find that about
$85\%$ of the flux is filtered out in the interferometric data.

It is difficult to derive a spectral index from the
continuum images because the different uv-coverages filter out
different amounts of flux.  However, we can measure fluxes
$S_{\rm{uv}}$ in the uv-plane. Ideally, one would select the same
regions in the uv-plane, but as this would reduce the amount of
available data even more, it is reasonable to compare the values for
matching baseline ranges (in units of $k\lambda$). We selected one
range of short baselines ($20-40\,k\lambda$, corresponding to spatial
scales between $10.7''$ and $5.4''$) and one range of longer baselines
($60-80\,k\lambda$, corresponding to spatial scales between $3.6''$
and $2.7''$) where there were sufficient data in both frequency bands:
the flux values are shown in Table \ref{para}. The 3.6\,cm flux is
only 0.9\,mJy \citep{sridha}, and assuming free-free emission its
contribution to the sub-(mm) observations is negligible.  Assuming a
power-law relation $S\propto \nu^{2+\beta}$ in the Rayleigh-Jeans
limit with the dust opacity index $\beta$, we find $\beta\sim 1.2$ for
short baselines corresponding to large spatial scales and $\beta\sim
0.5$ for large baselines corresponding to small spatial scales. These
values are lower than the canonical value of 2 \citep{hildebrand1983}
and decrease to small spatial scales. The exact values should be taken
with caution due to the large calibration uncertainty (estimated to be
within $25\%$), but the trend of a low $\beta$ decreasing with
decreasing spatial scales appears real. Without mapping the selected
sets of uv-data in the image-plane, we cannot determine whether the
smaller spatial scales correspond to the central core or to another
unresolved structure. However, the 850\,$\mu$m continuum image, which
is based on larger baselines, is more compact than the 1.3\,mm
emission (Fig. \ref{continuum}) indicating that the largest baselines
do trace the emission from the core center. A low value of $\beta$ has
also recently been found in another high-mass star-forming region
\citep{kumar2003}, and \citet{hogerheijde2000} report that $\beta$
decreases in L1489 from 1.5-2 in the envelope down to $\beta=1$ at the
inner peak. This trend may be due to grain growth within the central
core/disk or high optical depth.

Assuming optically thin dust emission at mm wavelength, we calculate
the mass and peak column density using the 1.3\,mm data following the
procedure outlined for the single-dish dust continuum data by
\citet{beuther2002a}. We use a grain emissivity index $\beta = 1$ and
a temperature typical for hot-core-like sources of 100\,K. The central
core mass then is $\sim 38$\,M$_{\odot}$ and the peak column density
$\sim 8\times 10^{23}$ cm$^{-2}$, corresponding to a visual extinction
$A_{\rm{v}}\rm{[mag]} = N_{\rm{H_2}}/(10^{21}\rm{cm}^{-2}) \sim
800$. As discussed by \citet{beuther2002a}, the errors are dominated
by systematics, e.g., exact knowledge of $\beta$ or the
temperature. We estimate the masses and column densities to be correct
within a factor of 5.

\subsection{The outflow in SiO}

Figure \ref{channel} presents a channel map of SiO(5--4),
and we find blue and red emission (systemic velocity
34.9\,km\,s$^{-1}$, \S \ref{hcooch3}) north of the continuum core. The
SiO emission by itself allows an interpretation of a bipolar outflow
centered $2''-3''$ north of the core. However, we do not detect any
sub-mm continuum source there ($850\,\mu$m $1\sigma$ mass sensitivity
$\sim 1$\,M$_{\odot}$), and a massive outflow without a sub-mm
continuum source at the driving center is very unlikely. Therefore, we
favor the interpretation of an outflow emanating from the core toward
the north with a position-angle (P.A.)  of $\sim 20^{\circ}$. The fact
that we see blue and red SiO emission toward the northern lobe
indicates that the outflow axis is near the plane of the sky.
For the red lobe, we find an increase in velocity with distance from
the center of the core resembling the Hubble-law within molecular
outflows (e.g., \citealt{downes1999}).

Previous unpublished single-dish SiO(2--1) observations with $29''$
resolution also show a bipolar outflow in the north-south
direction. The main difference is that the southern part of the
SiO(2--1) outflow is not observed at the higher frequency and higher
spatial resolution with the SMA. We can attribute this difference to
several possible reasons. First, interferometers filter out the large
scale emission, and without additional short spacing information, only
the more compact jet-like emission is detected with the SMA. Second,
the excitation of the 5--4 transition is higher with respect to the
2--1 transition. While the SiO(2--1) line traces the lower temperature
gas, the SiO(5--4) line is sensitive to the more excited gas being
associated with the more collimated component of the
outflow. Furthermore, an asymmetry in the small-scale distribution of
the dense gas can also contribute to the differences of the
large-scale SiO(2--1) and small-scale SiO(5--4) observations.

\subsection{HCOOCH$_3$ tracing the disk?}
\label{hcooch3}

\begin{figure}[htb]
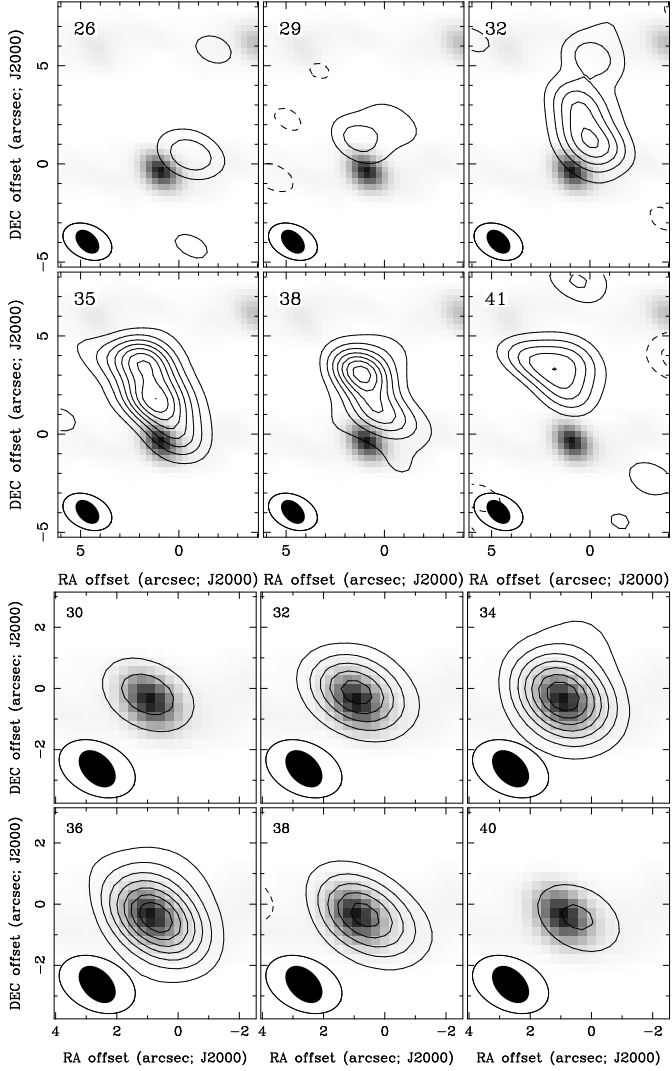
 
\includegraphics[angle=-90,width=8.8cm]{f2a.ps}\\
\includegraphics[angle=-90,width=8.8cm]{f2b.ps}
\caption{SiO(5--4) (top) and HCOOCH$_3$ (bottom) channel maps. The
grey-scale outlines the 850\,$\mu$m continuum emission. At the bottom
left of each panel, the filled ellipses shows the 850\,$\mu$m beam and
the unfilled ellipses the 1.3\,mm beam. LSR velocities (in
km\,s$^{-1}$) are given in the upper left corner of each panel. The
LSR velocities for HCOOCH$_3$ are calculated with respect to the line
component at 216.9662\,GHz, the systemic velocity is $\sim
34.9$\,km\,s$^{-1}$.
\label{channel}}
\end{figure}

\begin{figure}[htb] 
\includegraphics[angle=-90,width=8.8cm]{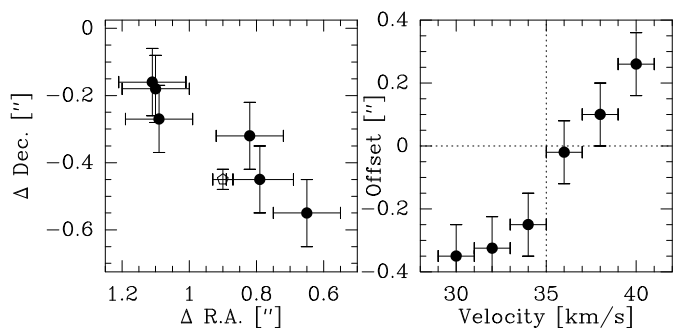}
\caption{
Presented are the
position-position (p-p) and position-velocity (p-v) diagrams of
HCOOCH$_3$. The offsets in the p-p plot were fitted in the uv-plane
for each velocity channel presented in Fig. \ref{channel}. The
positions delineate the P.A. of $55^{\circ}$, and the error bars
represent statistical errors of the fits. The open pentagon marks the
position of the 1.3\,mm continuum peak. The offsets in p-v diagram are
offsets from the continuum center position along the assumed disk plane
(P.A. $55^{\circ}$).
\label{pos_velo}}
\end{figure}

The high-density-gas tracing molecular HCOOCH$_3$(20--19) emission is
confined to the central core traced by the (sub-)mm continuum
(Fig. \ref{channel}). A Gaussian fit to the central spectrum results
in a systemic velocity of $\sim 34.9$\,km\,s$^{-1}$, and we observe a
velocity gradient across the core. It is possible to fit the peak
position of each spectral channel shown in Figure \ref{channel} to a
higher accuracy than the nominal spatial resolution, down to
0.5\,HPBW/(S/N), with S/N the signal-to-noise ratio
\citep{reid1988}. We performed these fits in the uv-plane to avoid
artifacts due to inverting and cleaning the data, the resulting
positions are shown in Figure \ref{pos_velo}. The derived velocity
gradient follows a P.A. of $\sim 55^{\circ}(\pm 10^{\circ})$ from blue
to red velocities. The separation of the most blue and most red
positions is $\sim 0.6''$, corresponding to $\sim 2200$\,AU. Figure
\ref{pos_velo} also presents a position-velocity diagram for the
fitted channels at the P.A. of $55^{\circ}$ confirming the
space-velocity shift. This velocity gradient is neither in the
direction of the SiO outflow nor directly perpendicular to it.

A velocity gradient solely due to a rotating disk would be
perpendicular to the outflow. The SiO outflow is near the plane of the
sky complicating an accurate determination of its P.A., and the
outflow might also be affected by precession (e.g., IRAS\,20126+4104,
\citealt{shepherd2000}). Regarding the HCOOCH$_3$ emission, it is
unlikely that this high-density tracer depicts a different
outflow which would remain undetected by the SiO. Therefore, we regard
neither the partially uncertain outflow P.A. nor a second outflow
sufficient to explain the difference in P.A. between the SiO and the
HCOOCH$_3$ emission. However, as shown in the accompanying paper
(Beuther et al., this volume), most molecular line data are influenced
by the molecular outflow, and it is likely that HCOOCH$_3$ is also
affected by the outflow. Furthermore, \citet{ohashi1996} have
shown that infall can influence the dense-gas velocity signature as
well. Assuming that the inner core contains a rotating disk, which has
an orientation perpendicular to the outflow, and that the HCOOCH$_3$
velocity signature is influenced by the disk, the outflow and infall,
it is plausible that the observed axis of the HCOOCH$_3$ velocity
gradient is offset from the expected disk orientation.  We do not
observe the inner region of the disk where the outflow/jet is
accelerated, but we may have detected the outer parts of the
accretion-disk.

Assuming equilibrium between centrifugal and gravitational forces at
the outer radius of the disk, we calculate the dynamical mass enclosed
within the central $0.6''$ to be $\sim 22/\rm{sin}^2i$\,M$_{\odot}$
(with $i$ the unknown inclination angle between the disk plane and
the line of sight, and an HCOOCH$_3$ velocity shift\footnote{The
full-width zero intensity is 12\,km\,s$^{-1}$. Taking into account the
4 dominant HCOOCH$_3$ line components spanning 3.5\,km\,s$^{-1}$ (\S
\ref{obs}), the HCOOCH$_3$ velocity shift is $(12-3.5)/2=4.25$.} of
4.25\,km\,s$^{-1}$), of the order of the core mass derived from the
dust emission (Table \ref{para}).  Thus, in contrast to T Tauri
systems where the disk mass is negligible compared with the
protostellar mass, in IRAS\,18089-1732 a considerable fraction of the
enclosed mass seems to be part of a large accretion disk and/or
rotating envelope. This result fits the picture that IRAS\,18089-1732
is in a very young evolutionary state, with the central (proto)star
still accreting material from the surrounding core/disk.

\section{Discussion}

The combined SiO(5--4) and HCOOCH$_3$(20--19) observations toward
IRAS\,18089-1732 support a massive star formation scenario where
high-mass stars form in a similar fashion as their low-mass
counterparts, i.e., via disk accretion accompanied by collimated
jets/outflows. The HCOOCH$_3$ observations still barely resolve the
disk/envelope structure and the interpretation is not entirely
unambiguous, but the data indicate rotation which might stem at least
partly from an accretion disk. Higher resolution observations in
different disk tracers are needed to investigate the disk/envelope
conditions in more detail. Furthermore, the continuum data indicate a
lower $\beta$ than the standard value of 2, and we observe a
decreasing $\beta$ with decreasing spatial scales. As nearly all the
gas observed in dust emission takes part in the assumed dynamical
rotation observed in HCOOCH$_3$ this low $\beta$ might be due to grain
growth or high opacity within a disk-like structure (e.g.,
\citealt{beckwith2000}).

\acknowledgments{It's a pleasure to thank the referee Riccardo
Cesaroni for his comprehensive comments improving the quality of
the paper. We also like to thank a lot the whole SMA staff for making
this instrument possible!  H.B. acknowledges financial support by the
Emmy-Noether-Program of the Deutsche Forschungsgemeinschaft (DFG,
grant BE2578/1).}



\begin{thebibliography}{17}
\expandafter\ifx\csname natexlab\endcsname\relax\def\natexlab#1{#1}\fi

\bibitem[{{Beckwith} {et~al.}(2000){Beckwith}, {Henning}, \&
  {Nakagawa}}]{beckwith2000}
{Beckwith}, S.~V.~W., {Henning}, T., \& {Nakagawa}, Y. 2000, Protostars and
  Planets IV, 533

\bibitem[{{Beuther} {et~al.}(2002{\natexlab{a}}){Beuther}, {Schilke}, {Menten},
  {Motte}, {Sridharan}, \& {Wyrowski}}]{beuther2002a}
{Beuther}, H., {Schilke}, P., {Menten}, K.~M., {et~al.} 2002{\natexlab{a}},
  \apj, 566, 945

\bibitem[{{Beuther} {et~al.}(2002{\natexlab{b}}){Beuther}, {Schilke},
  {Sridharan}, {Menten}, {Walmsley}, \& {Wyrowski}}]{beuther2002b}
{Beuther}, H., {Schilke}, P., {Sridharan}, T.~K., {et~al.} 2002{\natexlab{b}},
  \aap, 383, 892

\bibitem[{{Beuther} {et~al.}(2003){Beuther}, {Schilke}, \&
  {Stanke}}]{beuther2003a}
{Beuther}, H., {Schilke}, P., \& {Stanke}, T. 2003, \aap, 408, 601

\bibitem[{{Beuther} {et~al.}(2002{\natexlab{c}}){Beuther}, {Walsh}, {Schilke},
  {Sridharan}, {Menten}, \& {Wyrowski}}]{beuther2002c}
{Beuther}, H., {Walsh}, A., {Schilke}, P., {et~al.} 2002{\natexlab{c}}, \aap,
  390, 289

\bibitem[{{Cesaroni} {et~al.}(1999){Cesaroni}, {Felli}, {Jenness}, {Neri},
  {Olmi}, {Robberto}, {Testi}, \& {Walmsley}}]{cesaroni1999}
{Cesaroni}, R., {Felli}, M., {Jenness}, T., {et~al.} 1999, \aap, 345, 949

\bibitem[{{Churchwell}(2002)}]{churchwell2002}
{Churchwell}, E. 2002, \araa, 40, 27

\bibitem[{{Downes} \& {Ray}(1999)}]{downes1999}
{Downes}, T.~P. \& {Ray}, T.~P. 1999, \aap, 345, 977

\bibitem[{{Hildebrand}(1983)}]{hildebrand1983}
{Hildebrand}, R.~H. 1983, \qjras, 24, 267

\bibitem[{{Hogerheijde} \& {Sandell}(2000)}]{hogerheijde2000}
{Hogerheijde}, M.~R. \& {Sandell}, G. 2000, \apj, 534, 880

\bibitem[{{Kumar} {et~al.}(2003){Kumar}, {Fernandez}, {Hunter}, {Davis}, \&
  {Kurtz}}]{kumar2003}
{Kumar}, M.~S.~N., {Fernandez}, A.~J.~L., {Hunter}, T.~R., {Davis}, C.~J., \&
  {Kurtz}, S. 2003, \apj ~submitted

\bibitem[{{Ohashi} {et~al.}(1996){Ohashi}, {Hayashi}, {Ho}, {Momose}, \&
  {Hirano}}]{ohashi1996}
{Ohashi}, N., {Hayashi}, M., {Ho}, P.~T.~P., {Momose}, M., \& {Hirano}, N.
  1996, \apj, 466, 957

\bibitem[{{Reid} {et~al.}(1988){Reid}, {Schneps}, {Moran}, {Gwinn}, {Genzel},
  {Downes}, \& {Roennaeng}}]{reid1988}
{Reid}, M.~J., {Schneps}, M.~H., {Moran}, J.~M., {et~al.} 1988, \apj, 330, 809

\bibitem[{{Shepherd} {et~al.}(2001){Shepherd}, {Claussen}, \&
  {Kurtz}}]{shepherd2001}
{Shepherd}, D.~S., {Claussen}, M.~J., \& {Kurtz}, S.~E. 2001, Science, 292,
  1513

\bibitem[{{Shepherd} {et~al.}(2000){Shepherd}, {Yu}, {Bally}, \&
  {Testi}}]{shepherd2000}
{Shepherd}, D.~S., {Yu}, K.~C., {Bally}, J., \& {Testi}, L. 2000, \apj, 535,
  833

\bibitem[{{Sridharan} {et~al.}(2002){Sridharan}, {Beuther}, {Schilke},
  {Menten}, \& {Wyrowski}}]{sridha}
{Sridharan}, T.~K., {Beuther}, H., {Schilke}, P., {Menten}, K.~M., \&
  {Wyrowski}, F. 2002, \apj, 566, 931

\bibitem[{{Zhang} {et~al.}(1998){Zhang}, {Hunter}, \& {Sridharan}}]{zhang1998}
{Zhang}, Q., {Hunter}, T.~R., \& {Sridharan}, T.~K. 1998, \apjl, 505, L151

\end{thebibliography}


\begin{deluxetable}{lrrr}
\tablecaption{(Sub-)millimeter continuum data \label{para}}
\tablewidth{0pt}
\tablehead{
\colhead{} & \colhead{1.3mm$^a$} & \colhead{850$\mu$m$^a$} & \colhead{$\beta ^a$}
}
\startdata
$S_{\rm{peak}}$[mJy]& 690 & 1410 & \\
$S_{\rm{int}}^b$[mJy] & 1360& 1900 & \\
$S_{\rm{uv}}(20-40k\lambda)$[mJy] & 600 & 2870 & 1.2 \\
$S_{\rm{uv}}(60-80k\lambda)$[mJy] & 360 & 1240 & 0.5 \\
M[M$_{\odot}$]      & 38 &  & \\
N[$10^{23}$cm$^{-2}$] & 8.2 & & \\
\enddata
\tablenotetext{a}{\footnotesize Uncertainties are discussed in the main text.}
\tablenotetext{b}{\footnotesize $S_{\rm{int}}$ is measured within the $2\sigma$ contours levels}
\end{deluxetable}

\clearpage

\end{document}